\documentclass[10pt,conference]{IEEEtran}
\usepackage[T1]{fontenc}
\usepackage{amsmath,amssymb}
\usepackage{array,booktabs,tabularx}
\usepackage{xspace,cite,url}
\usepackage{pgfplots}
\usepackage{balance}
\usepgfplotslibrary{groupplots}
\usetikzlibrary{patterns,arrows.meta,positioning,calc}
\pgfplotsset{compat=1.18}
\usepackage[hidelinks]{hyperref}
\newcommand{\emac}{EmaC\xspace}
\newcommand{\compslo}{CompositeSLO\xspace}

\newcolumntype{Y}{>{\raggedright\arraybackslash}X}
\definecolor{emacblue}{RGB}{20,97,123}
\definecolor{emacorange}{RGB}{182,93,25}

\begin{document}
\title{Emergence-as-Code as a Foundation for Reliable Self-Governance}
\author{
  \IEEEauthorblockN{
    Anatoly A. Krasnovsky
    \IEEEauthorrefmark{1}\IEEEauthorrefmark{2}
  }
  \IEEEauthorblockA{
    \IEEEauthorrefmark{1}Innopolis University, Innopolis, Russia\\
    \IEEEauthorrefmark{2}MB3R Lab, Innopolis, Russia\\
  }
}
\maketitle

\begin{abstract}
Local adaptations can preserve component health while changing whether a
system meets its reliability requirements. The system-level consequence
depends on the interaction affected and its role in the user journey.
Emergence-as-Code (\emac) makes this relationship part of an executable,
evidence-reconciled CompositeSLO assessment. A declared journey obligation
persists while discovery maintains a hypothesis of its runtime realization.
An inferred operator-to-role binding selects the measurements used in the
calculation; its evidential status determines whether the binding supports a
numerical assessment. Each result retains the obligation and versioned
hypothesis as a basis for governance decisions. A PetClinic proof of concept
recovered the exact operator-state/edge-binding change in all 20 randomized
treatments, with no false change in 20 controls. EmaC and a manual dynamic
composite matched disjoint semantic outcomes in all 40 conditions. A model
with frozen suppression knowledge overestimated required-history availability
by 0.01, placing treatments above a 0.995 target despite observed availability
of 0.99. Ambiguous and contradictory evidence produced
\texttt{UNASSESSABLE}. The research programme evaluates sustained binding
maintenance, calibrated uncertainty, and the decision value of this
requirement-level account of local adaptation.
\end{abstract}

\begin{IEEEkeywords}
emergence, self-governance, self-adaptation,
model discovery, composite SLOs, uncertainty
\end{IEEEkeywords}

\section{Emergent Reliability and Self-Governance}
\label{sec:introduction}

Self-adaptive systems change their operation to meet declared
goals~\cite{kephart2003autonomic,delemos2013roadmap}. Bencomo, Welsh, Sawyer,
and Whittle connected emergent adaptive behavior to the need for explanations
in terms of requirement satisfaction~\cite{bencomo2012selfexplanation}.
Local adaptation creates a corresponding reliability problem: a circuit
breaker can protect a service while suppressing an interaction required by a
user journey. Component health and successful HTTP responses then leave the
system-level consequence undetermined. Assessment requires the current
relationship between local adaptation and the journey requirement.

In our PetClinic experiment, the owner-details gateway called Customers for
owner/pet information and Visits for history. An OPEN circuit breaker on one
gateway replica suppressed calls to the healthy Visits service. The gateway
returned HTTP-successful responses without history. These responses violated
the \emph{owner-history} requirement and satisfied \emph{owner-only}, which
allowed the omission. The same operator state and local indicators therefore
had different consequences for the two journeys. Interpreting the rejection
count required identifying the affected interaction and its declared role.

Self-governance means autonomous operation under human-declared objectives
and policies specifying permitted actions and evidence requirements.
The research question is whether maintaining the relationship between a
requirement and its changing realization improves autonomous decisions.

\paragraph*{Scientific foundation and proposed advance}
\emph{Emergence-as-Code} (\emac) develops requirement-oriented
self-explanation~\cite{bencomo2012selfexplanation,welsh2014selfexplanation}
into an executable basis for assessing the reliability consequences of local
adaptation. \emph{Emergence} denotes system-level behavior arising from
component interactions and local adaptation. Whether that behavior satisfies
a journey obligation depends on the current composition and the declared
success semantics. Building on requirements reflection~\cite{bencomo2010requirementsreflection},
runtime models and assurance~\cite{blair2009modelsruntime,cheng2014runtimeassurance},
and evidence-derived executable graphs~\cite{krasnovsky2026modeldiscovery},
we propose \emph{evidence-reconciled CompositeSLO assessment}. The proposed
advance is an executable assessment dependency: an application-specific
relationship between local adaptation and a persistent obligation is inferred
from observations and checked against declared semantics
(Sections~\ref{sec:proposal}--\ref{sec:instance}). Its accepted operator-to-role
binding selects the measurements entering the calculation; its evidential
status determines whether the binding supports a numerical assessment.
Each result retains the obligation and versioned realization hypothesis for
interpretation under a governance policy. Section~\ref{sec:evaluation}
evaluates the circuit-breaker instance, Section~\ref{sec:positioning} compares
the underlying relationships with prior approaches, and
Section~\ref{sec:agenda} sets out the tracking and decision studies.

\paragraph*{Contributions}
\textbf{C1} defines the obligation--realization--evidence relationship as an
executable assessment unit for self-governance.
\textbf{C2} instantiates it with operator/instance/edge inference, semantic
reconciliation, and a version-linked estimate or explicit inability to assess.
\textbf{C3} provides controlled evidence that discovery recovered the
application binding supplied to a manual dynamic composite, connecting
identification with accurate journey assessment and evidence-dependent refusal.

\begin{figure*}[t]
\centering
    \includegraphics[width=0.75\textwidth]{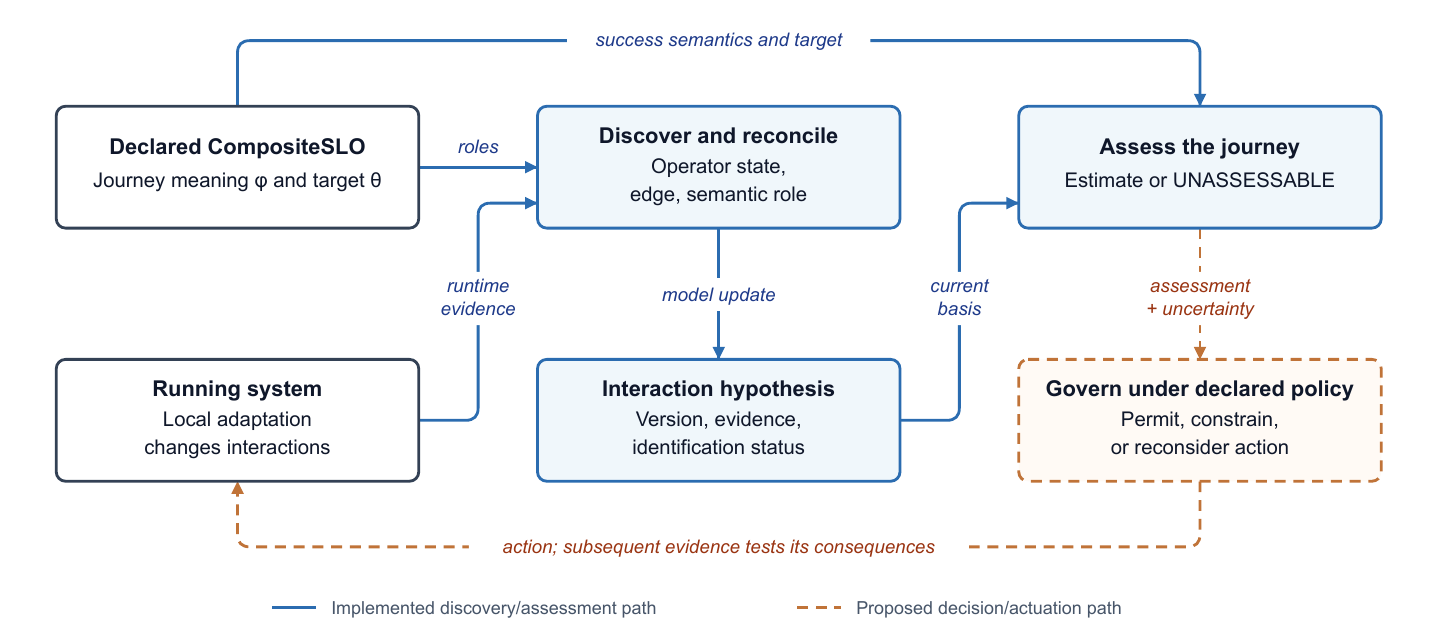}
\caption{EmaC connects a persistent CompositeSLO to a maintained hypothesis of
its runtime realization. The solid path implements discovery, reconciliation,
and assessment. The dashed path uses the assessment under a governance policy
and tests subsequent action effects; it is the closed-loop research programme.}
\label{fig:concept}
\end{figure*}

\section{Executable Obligations and Runtime Realizations}
\label{sec:proposal}

\subsection{A persistent requirement over changing interactions}

A \compslo expresses a journey's reliability requirement as code.
Conceptually, $J=(e,\varphi,\theta)$ specifies an entrypoint $e$, success
semantics $\varphi$ over declared interaction roles and supported composition
operators, and a target $\theta$. The requirement states the behavior that
must hold; the runtime model records the current realization. Suppressing an
interaction changes that realization while its obligation remains in $J$.
In this paper, journey availability is the fraction of relevant requests that
satisfy the declared success semantics.

The two PetClinic contracts use the owner-details endpoint and
$\theta=0.995$. Owner-history requires Visits history; owner-only permits the
empty-history fallback. A declared role predicate identifies the history
branch. The concrete runtime edge and operator that realize this role are
discovery outputs. Thus the contract supplies the meaning of success, and
observations supply the application-specific binding used to assess it.

The prototype stores this contract in versioned JSON. The role predicate
contains \texttt{"operationContains": ["/pets/visits"]}; the fields
\texttt{suppressedInteractionRole} and \texttt{fallbackSatisfiesJourney}
identify the role and specify whether fallback satisfies the journey.
A specialized compiler combines these declarations with discovered operator
behavior. A composition language covering multiple interacting operators
belongs to the research programme in Section~\ref{sec:agenda}.

This separation has an operational consequence. Removing an unobserved edge
from a runtime graph cannot establish that the requirement associated with it
has disappeared. Discovery must retain the role as an assessment obligation
and determine whether the evidence supports a current realization. A reported
value therefore depends on both the declared semantics and the supported
interpretation of runtime observations.

\subsection{The assessment unit and its use by a policy}

Figure~\ref{fig:concept} shows how this relationship enters the proposed loop.
For evidence $E_t$, previous model $M_{t-1}$, and adapter catalog $C_A$,
\begin{equation}
\begin{aligned}
 (\widetilde\Delta_t,\widetilde z_t)&:=\operatorname{discover}(E_t,M_{t-1};C_A),\\
 (M_t,z_t)&:=\operatorname{reconcile}(M_{t-1},\widetilde\Delta_t,\widetilde z_t),\\
 \mathcal A_{J,t}&:=(J,M_t,z_t,\operatorname{assess}(J,M_t,z_t)).
\end{aligned}
\label{eq:assessment}
\end{equation}
Discovery proposes a typed model change and an identification status.
Reconciliation checks the proposal against declared roles and model structure.
$M_t$ records the resulting realization hypothesis, including operator bindings
and parameters. Its status $z_t$ distinguishes identified, unresolved, and
contradictory knowledge. The implemented compiler emits a conditional point
estimate or \texttt{UNASSESSABLE} when a required binding lacks support.

The assessment unit $\mathcal A_{J,t}$ carries the requirement together with
the model version and evidence status that gave the numerical result its
meaning. This retained relationship is the systemic contribution of EmaC.
Operator-to-role inference supplies a concrete mechanism for constructing it.
The same realization can support assessments of several journeys with
different success semantics, as the two PetClinic contracts demonstrate.

A governance policy interprets $\mathcal A_{J,t}$ under permitted actions,
risk tolerances, and recovery rules. A supported estimate below target can
prompt a configuration change; an unresolved realization can prompt further
observation or defer a model-dependent action. Action selection also requires
models of action effects and subsequent observations of their consequences.
The implemented path supplies and preserves the assessment; F3 studies its
use in autonomous decisions.

\section{Inferring and Assessing an Operator-to-Role Binding}
\label{sec:instance}

\subsection{Declared knowledge and runtime inference}

Table~\ref{tab:contract} defines the inputs and inferred facts. A generic
Resilience4j adapter supplies metric syntax and circuit-breaker
semantics~\cite{resilience4jCircuitBreaker}. Metrics and OpenTelemetry
traces~\cite{opentelemetrySpec} provide observations of the application
operator, replica, affected edge, state, and routing fraction. Bootstrap traces
record normally executed edges. Current traces expose missing executions;
per-instance metrics record operator state and decision counts. Their shared
instance identity connects these observations in a binding hypothesis.

\begin{table}[t]
\caption{Declared semantics, inferred realization, and evaluation-only information.}
\label{tab:contract}
\footnotesize
\begin{tabularx}{\columnwidth}{@{}>{\raggedright\arraybackslash}p{.25\columnwidth}Y@{}}
\toprule
Information & Source and role \\
\midrule
Journey obligation & Declared entrypoint, target, history-role predicate, and fallback success value $A_F$. \\
Operator vocabulary & Declared generic adapter semantics and counter syntax. \\
Runtime realization & Inferred operator, instance, state, affected edge, and decision-weighted parameters. \\
Effective hypothesis & Reconciled fields, identification status, evidence references, and version lineage. \\
Ground truth & Hidden intervention and semantic outcome oracle; evaluation after freeze. \\
\bottomrule
\end{tabularx}
\end{table}

Let $N$ be completed evidence-window requests and $D_{io}$ the decision-counter
difference for operator name $o$ on instance $i$. The adapter selects a unique
name satisfying
\begin{equation}
 \left|\sum_iD_{io}-N\right|\leq\tau(N),\qquad
 \tau(n)=\max\{1,\operatorname{round}(0.01n)\}.
\label{eq:selection}
\end{equation}
Rounding is to the nearest integer, with ties to even. Zero or multiple
matching names produce unresolved status. At $N=6{,}000$, the allowed
count difference is 60. Substantial failure before the decision point can
therefore prevent selection even when an operator is relevant to the journey.
This exposure condition is part of the adapter's identification domain.

For the selected operator and an instance with $r_i>0$ rejected decisions,
let $n_i$ be observed journey traces, $x_{ie}$ executions of edge $e$, and
$b_{ie}$ its bootstrap execution rate. Count-consistent candidates are
\begin{equation}
 C_i=\{e:b_{ie}\geq0.95\ \land\ |(n_i-x_{ie})-r_i|\leq\tau(n_i)\}.
\label{eq:binding}
\end{equation}
One candidate permits a proposed binding; multiple candidates yield
unresolved status and zero candidates yield contradictory status.
Reconciliation checks the proposal against bootstrap structure and the
history-role predicate. A unique wrong-role edge produces
\texttt{UNASSESSABLE}. The $0.95$ cutoff defines bootstrap eligibility, and
the unsuppressed branch must execute once per relevant request. At $n_i=60$,
the count tolerance is one. Identification is conditional on the adapter and
observed candidates; unmodeled causes can also produce matching counts.

At full coverage of the 60-request minority, $n_i=r_i=60$ and $x_{ie}=0$
for the suppressed history edge give residual zero in
Equation~\ref{eq:binding}. An edge executed on all 60 traces gives residual
60 and cannot match. Numerical agreement becomes a supported history binding
after the semantic-role check. Two eligible edges with the same matching
residual leave the choice unresolved. This sequence separates observed
suppression, inferred application identity, and declared journey meaning.

\subsection{From the inferred relationship to a journey estimate}

The accepted binding selects the operator counters entering the calculation.
Let $D=\sum_iD_i$, $P=\sum_iP_i$, $S=\sum_iS_i$, and $r=\sum_ir_i$ aggregate
decisions, permitted calls, successful permitted calls, and rejections.
For positive denominators,
\begin{equation}
 A_P=D/N,\qquad q=P/D,\qquad A_V=S/P.
\label{eq:parameters}
\end{equation}
$A_P$ represents reaching the decision point, $q$ the permitted fraction, and
$A_V$ conditional success of permitted calls. The compiler assigns fallback
value $A_F$ to both rejected and failed permitted calls:
\begin{equation}
 \widehat R_J=A_P\bigl[qA_V+(1-qA_V)A_F\bigr]
             =\frac{S+(D-S)A_F}{N}.
\label{eq:compiler}
\end{equation}
These factors are nested count ratios. The model treats a successful permitted
call as journey success, applies the declared fallback value to the other
$D-S$ decisions, and assigns no success to the $N-D$ requests outside the
decision point. These assumptions define the implemented composition.

The count form requires $N,D>0$, $0\leq S\leq P\leq D\leq N$, and $P+r=D$.
Inconsistent counts are contradictory; absent exposure is unresolved.
If all calls are rejected, $P=S=0$ leaves $A_V$ undefined, while the count
form still evaluates the fallback under a supported binding. Owner-history
sets $A_F=0$; owner-only sets $A_F=1$. The binding determines which measurements
apply, and the contract determines their journey-level consequence.
For these contracts, the count form gives
$\widehat R_{\mathrm{history}}=S/N$ and
$\widehat R_{\mathrm{owner}}=D/N$, with a difference of $(D-S)/N$.
Thus a shared set of local decision counts supports different assessments
according to the role's obligation. The semantic distinction remains fixed
while discovery supplies the current operator/interaction relationship.
Successful fallback counts toward owner-only satisfaction while omitted
history continues to count against owner-history satisfaction.

\subsection{Preserving the derivation}

The pipeline bootstraps $M_0$, proposes a typed delta, reconciles it, and
compiles the effective model. The catalog, bootstrap, candidate delta,
reconciliation, model, and estimate have content-derived versions checked
before use. A wrapper reconstructs the model from its source artifacts,
requires exact equality, and recomputes the terminal estimate before freeze.
The compiler reads the contract and effective model through a procedural
boundary on a shared filesystem. Unresolved or contradictory required bindings
yield a versioned model and \texttt{UNASSESSABLE}; integrity violations are
errors. These checks preserve the computation's derivation. The evaluation
checks binding identity against the hidden intervention and the estimate
against independently collected semantic outcomes.

\section{Proof of Concept: Binding, Evidence, and Consequence}
\label{sec:evaluation}

\subsection{Questions and controlled protocol}

The study evaluated \textbf{E1}, recovery of the injected operator-state/edge-binding
change and false control changes; \textbf{E2}, assessment behavior under
insufficient or conflicting evidence; and \textbf{E3}, numerical agreement
with held-out journey outcomes. Binding identity and assessment accuracy were
separate endpoints.

The artifact used a pinned AWS Application Signals version of Spring
PetClinic~\cite{awsPetclinicDemo}. A runtime overlay supplied instrumentation,
routing, fault injection, and isolation of unrelated demo components; business
source and the existing fallback remained unchanged. The companion
repository~\cite{emacPetclinicArtifact} records the experiment version and
workflow run and contains code, protocol, schemas, tests, and paired and
aggregate reports.

Twenty independently reset pairs executed control and treatment in seeded
random order. Two gateway replicas received fresh opaque identities. One was
randomly assigned the deterministic 1\% traffic minority; discovery received
neither mapping. Each condition bootstrapped from 200 fully traced requests
at 25 requests/s, balanced 100 per replica, froze $M_0$, and restarted the
gateways to discard breaker history while preserving identities. Measurement
then used 99:1 routing. Control remained CLOSED. Treatment preconditioning
used failed and rejected calls to leave the minority breaker OPEN. The
experiment disabled the injected fault and verified Visits health before
measurement; OPEN persisted through both windows.

A 6,000-request evidence window at 50 requests/s supplied metrics and traces.
The collector discarded response bodies. All model and assessment versions
froze before a disjoint 6,000-request window at the same rate. The evaluator
then used the hidden intervention and semantic outcome oracle. Measurement
checks required model-before-outcome ordering, exact counter boundaries,
at least 99\% trace coverage, matching run IDs, and paired local availability
within 0.1 percentage point for gateway, Customers, and Visits. All 20 pairs
passed, with zero exclusions or replacements. The reset pair was the replicate.

The protocol retained wrong bindings and \texttt{UNASSESSABLE} results in the
denominator. A baseline refusal preserved the EmaC result and outcome
collection. Completion required every scheduled secondary analysis, including
those recording method failures; all completed. Thus measurement validity,
discovery success, and report completion remained distinct checks.

\subsection{Baselines and evidence conditions}

The \emph{manual dynamic composite} received the operator/edge/fallback
mapping and used Equation~\ref{eq:compiler}. Matching it tested whether
observations supplied the application knowledge needed for assessment.
The \emph{frozen model} retained $q=1$ with measured $A_P,A_V$, isolating the
consequence of stale suppression knowledge. Metrics-only and traces-only
views separated operator observations from interaction observations.

Constructed ambiguity supplied two count-consistent edges; constructed
contradiction supplied none. Both passed through the same discovery,
reconciliation, and compilation path. Identity redaction removed the
cross-source instance key. These conditions replayed frozen pre-outcome
evidence and tested the response to particular losses of binding support.

Full fusion recovered the exact typed change in all 20 treatments and produced
no false delta in 20 controls (Table~\ref{tab:results}). Metrics identified
operator/state facts; traces identified the affected edge. Their shared
instance key supported the fused binding. Ambiguous and contradictory
packages produced \texttt{UNASSESSABLE} in all 20 replays of each kind.
Redacted identities left aggregate $q$ available and the instance binding
unresolved.

\begin{table}[t]
\caption{Binding recovery and evidence-dependent assessment. Numeric rows use 20 conditions or evidence packages.}
\label{tab:results}
\footnotesize
\begin{tabularx}{\columnwidth}{@{}>{\raggedright\arraybackslash}p{.31\columnwidth}Y@{}}
\toprule
Evidence condition & Result \\
\midrule
Full fusion, treatment & Exact operator state + edge binding: 20/20 \\
Full fusion, control & False delta: 0/20 \\
Metrics-only view & Operator/state; edge binding unresolved \\
Traces-only view & Edge; operator/state unresolved \\
Ambiguous evidence & \texttt{UNASSESSABLE}: 20/20; numeric output: 0 \\
Contradictory evidence & \texttt{UNASSESSABLE}: 20/20; numeric output: 0 \\
Identity-redacted view & Aggregate $q$; instance binding unresolved \\
\bottomrule
\end{tabularx}
\end{table}

\begin{figure*}[t]
\centering
    \includegraphics[width=0.9\textwidth]{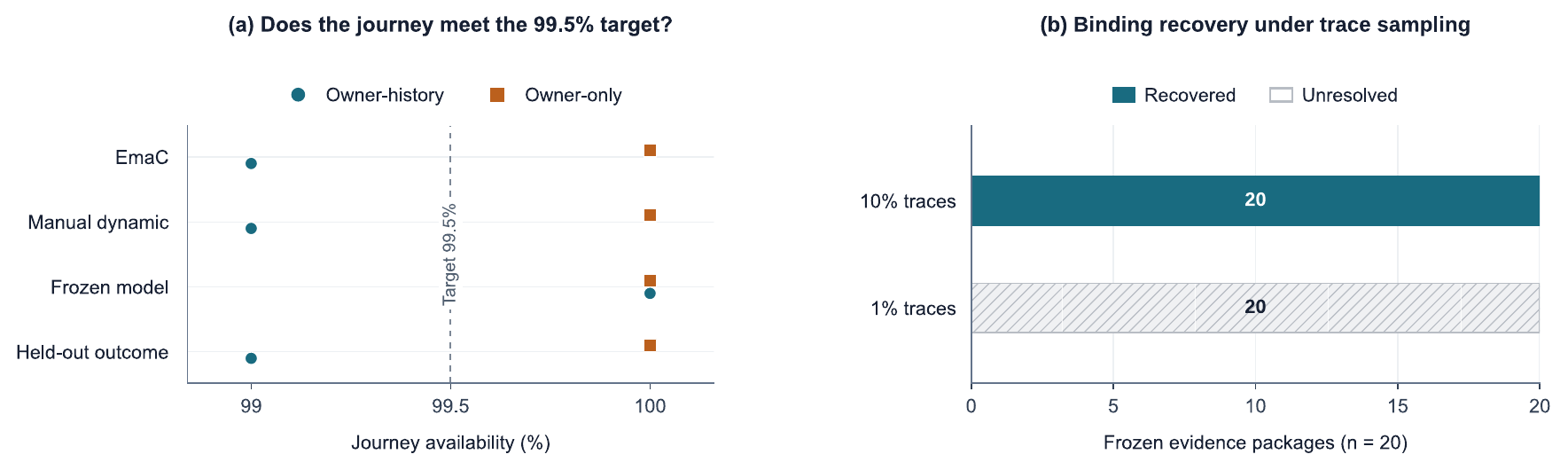}
\caption{Assessments and binding recovery. (a) Each treatment yielded 99\%
owner-history availability for EmaC, the manual dynamic composite, and the
held-out outcomes; the frozen model gave 100\%, above the 99.5\% target.
Owner-only remained at 100\%; both journeys were at 100\% in controls.
(b) Sampling replays of the 20 frozen evidence packages retained full metrics
and used Equation~\ref{eq:sampling}: 10\% traces gave 20 recoveries; 1\% gave
20 unresolved bindings. No sampled replay produced a false binding.}
\label{fig:results}
\end{figure*}

Sampling replays kept full metrics and hash-sampled the frozen traces. Their
separate rate heuristic required a single instance with rejected calls,
at least three sampled traces on that instance, and a unique
bootstrap-eligible edge satisfying
\begin{equation}
 \left|\frac{n_i^{(s)}-x_{ie}^{(s)}}{n_i^{(s)}}-\frac{r_i}{D_i}\right|\leq0.01.
\label{eq:sampling}
\end{equation}
Fewer traces or multiple rejected instances yielded unresolved status.
At 10\% sampling, all 20 packages remained recoverable; at 1\%, all 20 became
unresolved, with no false binding at either rate (Figure~\ref{fig:results}b).
The minority received 60 requests per window, giving expected sampled counts
of six and 0.6. Primary recovery used the count rule in
Equation~\ref{eq:binding}; these replays used Equation~\ref{eq:sampling}.

Mutation tests rejected changes to $q$ or the graph without matching versions;
a resealed wrong-role edge yielded \texttt{UNASSESSABLE}. All 27 automated
tests passed, covering absent or ambiguous operators, invalid counts,
all-rejected windows, retention of unsuccessful discovery, baseline refusal,
and incomplete secondary reports. For 20/20 successes, the two-sided 95\%
exact-binomial lower bound is 0.832. Zero false deltas in 20 controls gives
an upper bound of 0.168.

\subsection{Requirement-level consequences}

Paired local-availability differences were 0.000 percentage points for all
three services. Owner-history satisfied the requirement in 5,940/6,000
treatment outcomes; owner-only satisfied it in all 6,000. The treatment had
$A_P=A_V=1$ and $q=0.99$. Substituting the two fallback declarations into
Equation~\ref{eq:compiler} gave 0.99 and 1.00, respectively.

EmaC and the manual dynamic composite produced numerical assessments for both
journeys in all 40 conditions, with zero error against the oracle and no
abstentions. The frozen model was correct in controls and for owner-only.
It overestimated owner-history by 0.01 in every treatment, placing its 1.00
estimate above the 0.995 target while the observed value was below it
(Figure~\ref{fig:results}a).

The experiment connected a recovered local operator/interaction relation to
two different system-level obligations. Its demonstrated result is recovery
of application knowledge that the manual composite received as input.
A counter monitor supplied the same binding and fallback semantics would
compute the same values. Routing and suppression were deterministic across
the evidence and outcome windows, so the study evaluated the implemented
path under repeated controlled execution. The sustained-maintenance and
decision comparisons follow in Section~\ref{sec:agenda}.

\section{Relationship to Existing Approaches}
\label{sec:positioning}

The relevant distinction concerns the relationship connecting a requirement
to its current realization: how a method obtains it, updates it, and uses it
in assessment. Table~\ref{tab:positioning} compares the closest mechanisms.
EmaC combines a persistent obligation with an inferred operator-to-role
binding whose evidential status controls its use in quantitative assessment.

\begin{table}[t]
\caption{Relationships maintained or evaluated by nearby approaches and the EmaC instance.}
\label{tab:positioning}
\footnotesize
\begin{tabularx}{\columnwidth}{@{}>{\raggedright\arraybackslash}p{.29\columnwidth}Y@{}}
\toprule
Approach & Relationship and use in analysis \\
\midrule
Model Discovery~\cite{krasnovsky2026modeldiscovery}
& Operational artifacts yield executable interaction graphs with SLO metadata. \\
Goal-based adaptation~\cite{cailliau2017obstacles,chen2014uncertainty}
& Goal/obstacle refinements propagate monitored probabilities; functional alternatives have updated quality contributions. \\
Architectural maintenance~\cite{heinrich2020iobserve,mazkatli2025cipm}
& Implementation correspondences and observations update architecture, call structure, and quantitative parameters. \\
ENTRUST~\cite{calinescu2018assurance}
& Runtime evidence completes application-specific partial models for requirement verification and assurance cases. \\
EmaC instance
& A declared role and operator vocabulary constrain an inferred operator/edge binding; its evidence status governs counter-based assessment. \\
\bottomrule
\end{tabularx}
\end{table}

\paragraph*{Emergence, requirements, and adaptive control}
Bencomo et al. motivated explaining emergent behavior through requirement
satisfaction~\cite{bencomo2012selfexplanation}; Welsh et al. derived
self-explanations from runtime goal models and encoded them in a domain-specific
language~\cite{welsh2014selfexplanation}. Requirements reflection makes
requirements runtime entities~\cite{bencomo2010requirementsreflection}.
Requirements@Runtime combines requirements and adaptation capabilities and
propagates their changes to executing applications~\cite{pasquale2011requirementsruntime}.
Awareness requirements express satisfaction-related conditions on the feedback
loop~\cite{souza2011awareness}, and REAssuRE revisits design-time assumptions
about goal realization~\cite{welsh2011reassure}.

DYNAMICO coordinates changes in control objectives, adaptation mechanisms,
and monitoring relevance~\cite{villegas2013dynamico}. EUREMA and ActivFORMS
provide executable feedback-loop models and formally founded adaptation
infrastructure~\cite{vogel2012eurema,weyns2023activforms}. These approaches
establish system-level integration of requirements and adaptation. EmaC's
proposed assessment unit specifies the knowledge supplied to such control:
which local operator affects a persistent role, how this changes the
calculation, and whether the inferred relationship has evidential support.

\paragraph*{Quantitative goals and assurance}
Cailliau and van Lamsweerde propagate monitored obstacle probabilities through
goal/obstacle refinement trees and select countermeasures against required
satisfaction rates~\cite{cailliau2017obstacles}. Chen et al. update quantitative
contributions from functional alternatives to quality goals and handle
uncertain preferences and action timing~\cite{chen2014uncertainty}. KAMI and
adaptive model learning update quantitative behavior from
execution~\cite{epifani2009kami,calinescu2014adaptivelearning}; QoSMOS and
continuous assurance connect such models to adaptation and requirement
verification~\cite{calinescu2011qosmos,filieri2012continuous}.

ENTRUST permits incomplete models, including missing structural parts, and
combines runtime verification with dynamic assurance
cases~\cite{calinescu2018assurance}. EmaC provides a domain-specific inference
and compilation procedure for an unknown operator-to-role relationship.
Matching rejection counts to missing required interactions determines the
application binding used in the reliability calculation. The supported
binding and its categorical identification status travel with the result.

\paragraph*{Model discovery and architectural correspondence}
Model Discovery and Graph Simulation derives executable graphs from
operational artifacts and connects them with SLO analysis~\cite{krasnovsky2026modeldiscovery}.
Rainbow provides architecture-based adaptation~\cite{garlan2004rainbow};
incremental synchronization and bidirectional transformations maintain
runtime model consistency~\cite{vogel2010synchronization,song2011runtimearchitecture}.
Perera et al. manage and visualize runtime artifact traceability within a
self-adaptive framework~\cite{perera2015traceability}. iObserve combines
observations with implementation--architecture
correspondence~\cite{heinrich2020iobserve}. CIPM recovers service-call
structure and maintains calibrated architectural performance
models~\cite{mazkatli2025cipm}.

EmaC extends this model-maintenance perspective to the interpretation of
suppressed interactions under a persistent obligation. The missing execution
remains relevant to assessment even when it contributes no current call edge.
Runtime-model provenance and uncertainty research supply techniques for
recording the evidence and conditions of this
interpretation~\cite{reynolds2020provenance,hezavehi2021uncertainty}.

\paragraph*{SLO composition and direct observation}
SRE practices define user-relevant indicators~\cite{beyer2018sreworkbook};
OpenSLO expresses multiple objectives and composite weights~\cite{opensloSpec},
and ParSLO allocates latency targets over microservice
structures~\cite{mirhosseini2021parslo}. Resiliency-pattern analysis models
retries and circuit breakers with continuous-time Markov chains to evaluate
tradeoffs~\cite{mendonca2020patterns}. EmaC binds observed operator execution
to the role whose success semantics determine the assessment. A content-aware
SLI or role-aware trace monitor can directly measure a journey's current
outcomes. The additional result supplied by EmaC is the supported
operator/interaction relationship behind the calculation; F3 evaluates its
use in selecting actions at matched observation budgets.

\section{Scope and Research Programme}
\label{sec:agenda}

The proof of concept covers one application, circuit-breaker adapter, affected
edge, and point-valued compiler. Journey semantics and generic operator
vocabulary are declared. The count rule requires compatible request exposure
and trace coverage; retries, cancellations, and asymmetric trace loss can
break that correspondence. Concurrent operators and optional branches can
produce several explanations. The sampling result concerns deterministic
suppression of the minority branch. The programme therefore evaluates binding
maintenance, uncertainty, and decision outcomes as separate properties of the
same assessment unit. Aggregate trace coverage also leaves the distribution
of missing observations across replicas unspecified; per-instance exposure
is therefore a distinct factor in the planned sensitivity analysis.

\paragraph*{F1---Maintain the realization through adaptation}
Randomized no-reset blocks will vary retries, fallback, rate limiting, and
routing across multiple instances and edges. Hidden interventions will be
revealed after each model freezes. Comparators are a structure-only updater,
an always-select-best-match variant, a dynamic counter monitor with a fixed
binding, and a role-aware trace monitor using the same evidence. The study
will measure typed-change recovery, false bindings, unresolved fraction, and
detection delay separately from assessment error. State changes within a
binding and changes to the binding itself will be scored separately. Control
blocks will hold the relationship fixed; recovery delay will count windows
from a hidden transition until the first supported correct binding.

Sensitivity analysis will cross count tolerances of 0.5\%, 1\%, 2\%, and 5\%
with bootstrap cutoffs, minority exposure, optional branches, asymmetric trace
loss, and distractor edges. Replay will identify failure regimes for subsequent
live tests. This sequence tests when updating the operator/role relationship
preserves a useful assessment and when the evidence requires withholding it.
Wrong and unresolved bindings remain in the denominator; independently defined
instrumentation failures determine measurement validity.

\paragraph*{F2---Calibrate requirement-level uncertainty}
Experiments will cross trace sampling, identity loss, nonstationary traffic,
and common-cause failures. Bounds will propagate both count uncertainty and
variation across admissible realization hypotheses. Controlled regimes with
known journey probabilities will test interval coverage and width; held-out
semantic outcomes will test shared model assumptions. Comparisons will report
target-side errors and unresolved cost at matched telemetry budgets. Assessment
coverage will accompany interval coverage so that the cost of withholding
estimates remains visible. Model error also includes admissible-hypothesis
sets that exclude the actual realization: an interval over such a set can be
narrow while misrepresenting the journey. The study will distinguish this
identification error from uncertainty within a supported hypothesis.

\paragraph*{F3---Measure the value for self-governance}
The assessed hypotheses will feed declared policies and action-effect models.
EmaC, frozen-model, manually maintained, and sparse content-aware-probe
controllers will face the same workload and fault schedules at matched
evidence budgets. Interventions with opposing effects across journeys will
test joint satisfaction of their CompositeSLOs. Records will include proposed
and executed actions, predicted effects, observed outcomes, and reasons for
deferral. The study will measure false promotion, false block, abstention,
decision delay, telemetry cost, and post-action objective violations.

The reset or temporal block remains the replicate. Studies will use paired
schedules, retain all attempts, and choose replication through pilot-based
precision or power analysis before confirmatory execution. Progression
requires correct binding updates with useful assessment coverage in F1,
nominal interval coverage and fewer target-side errors in F2, and improved
decisions at comparable observation and deferral costs in F3.

\section{Conclusion}

EmaC makes the current relationship between local adaptation and a persistent
system-level requirement part of an executable assessment. Discovery supplies
the application binding; reconciliation determines whether the evidence
supports its use; the result retains the requirement and versioned realization
hypothesis. In 20 PetClinic treatment/control pairs, this path recovered the
binding supplied to a manual composite and matched held-out journey outcomes.
The two contracts exposed different reliability consequences of the same
local suppression. The research programme tests whether maintaining this
relationship improves requirement-level assessment and autonomous decisions
as the realization changes.

\balance
\bibliographystyle{IEEEtran}
\bibliography{arxiv3}
\end{document}